\documentclass[12pt]{article}
\usepackage{graphicx}
\usepackage{latexsym}
\def\be{\begin{equation}}
\def\ee{\end{equation}}
\def\bea{\begin{eqnarray}}
\def\eea{\end{eqnarray}}
\begin{document}

%\begin{flushright}
%\today\\
%\end{flushright}

\begin{center}
{\Large \bf Curvature perturbations from ekpyrotic collapse with
multiple fields} \vskip 1cm

Kazuya Koyama$^{\dagger}$ \footnote{ E-mail: Kazuya.Koyama@port.ac.uk}
Shuntaro Mizuno$^{\ddagger}$ \footnote{ E-mail: mizuno@resceu.s.u-tokyo.ac.jp}
and
David Wands$^{\dagger}$ \footnote{E-mail: David.Wands@port.ac.uk }

\vskip 1cm $^{\dagger}$ Institute of Cosmology and Gravitation, Mercantile House,
University of Portsmouth, Portsmouth~PO1~2EG, United Kingdom \\
$^{\ddagger}$  Research Center for the Early Universe (RESCEU), School of
Science, University of Tokyo, 7-3-1 Hongo, Bunkyo, Tokyo~113-0033,
Japan
\end{center}

\begin{abstract}
A scale-invariant spectrum of isocurvature perturbations is generated
during collapse in the ekpyrotic scaling solution in models where
multiple fields have steep negative exponential potentials. The scale
invariance of the spectrum is realized by a tachyonic instability in
the isocurvature field. This instability drives the scaling solution
to the late time attractor that is the old ekpyrotic collapse
dominated by a single field. We show that the transition from the
scaling solution to the single field dominated ekpyrotic collapse
automatically converts the initial isocurvature perturbations about
the scaling solution to comoving curvature perturbations about the
late-time attractor. The final amplitude of the comoving curvature
perturbation is determined by the Hubble scale at the transition.
\end{abstract}

\section{Introduction}

The existence of an almost scale-invariant spectrum of primordial
curvature perturbations on large scales is one of the most important
observations that any model for the early universe should explain.
% DW2
An inflationary expansion in the very early universe is most
commonly assumed to achieve this,
but it is important to consider whether there is any alternative
model. In this paper we focus on the ekpyrotic scenario as an
alternative \cite{Khoury:2001wf} (see also \cite{Kallosh:2001ai,
Khoury:2001iy}). In the old ekpyrotic scenario, the large scale
perturbations are supposed to be generated during a collapse driven
by a single scalar field with a steep negative exponential
potential. It was shown that the Newtonian potential acquires a
scale-invariant spectrum, but the comoving curvature perturbation
has a steep blue spectrum \cite{Lyth:2001pf}. In this scenario we
need a mechanism to convert contraction to expansion, and for a
regular four-dimensional bounce, the scale-invariant Newtonian
potential is matched to the decaying mode in an expanding universe,
and the growing mode of curvature perturbations acquires a steep
blue spectrum \cite{Brandenberger:2001bs, Tsujikawa:2002qc}.
% DW ADDED TO NEXT SENTENCE
It has been suggested that this conclusion might be altered by
allowing a singular matching between collapse and expanion
\cite{Cartier:2003jz}, but the general rule that the comoving
curvature perturbation remains constant still holds
for adiabatic perturbations on large scales \cite{Copeland:2006tn}.
In a braneworld context, the conversion from contraction to
expansion might be accomplished by a collision of two branes where
one of extra-dimensions disappears \cite{Khoury:2001bz}. It was
argued that the scale-invariant Newtonian potential can be
transferred to the comoving curvature perturbations by this singular
bounce \cite{Tolley:2003nx}. However, without having a concrete
theory to describe the singularity, it is difficult to have a
definite conclusion on how perturbations pass through the
singularity.

Recently, there has been some progress in generating a scale
invariant spectrum for curvature perturbations in the ekpyrotic
scenario  \cite{Lehners:2007ac, Buchbinder:2007ad,
Creminelli:2007aq}, by considering non-adiabatic perturbations which
has been suggested previously by Ref.~\cite{Notari:2002yc}. In this
case we require two or more fields. If these have steep exponential
potentials then there exists a scaling solution where the energy
density of the fields grow at the same rate during collapse
\cite{Finelli:2002we, Guo:2003eu}. The isocurvature perturbations
then have a scale-invariant spectrum \cite{Finelli:2002we}. These
isocurvature perturbations can be converted to curvature
perturbations if there is a sharp turn in the trajectory in field
space \cite{Lehners:2007ac, Buchbinder:2007ad, Creminelli:2007aq}.
For example Ref.~\cite{Lehners:2007ac} considered a situation where
one of the fields changes its direction in field space, which
corresponds to a time when a negative tension brane is reflected by
a curvature singularity in the bulk, in the context of the heterotic
M-theory. Refs.~\cite{Buchbinder:2007ad, Creminelli:2007aq}
considered a regular bounce realized by a ghost condensate. One of
the fields exits the ekpyrotic phase and hits the transition to the
ghost condensate phase that creates a sharp turn in the trajectory
in field space and curvature perturbations can be generated
\cite{Buchbinder:2007ad}. It is still necessary to match curvature
perturbations in a contracting phase to those in an expanding
universe, but it is shown that the comoving curvature perturbation
is conserved on large scales resulting in an almost scale-invariant
spectrum observed today for a regular bounce like a ghost condensate
model \cite{Buchbinder:2007ad}.

The isocurvature perturbations behave like $\delta s \propto H$ on
large scales. As the Hubble parameter is rapidly increasing in a
collapsing universe, this signals an instability. In fact it is easy
to see that we always require an instability of this form in order
to generate a scale-invariant spectrum with a canonical scalar field
during collapse\footnote{The only way to produce a scale-invariant
spectrum without the presence of an instability seems to be due to
non-canonical kinetic terms, as in the case of axion-type fields
which can acquire scale-invariant perturbation spectra while
remaining massless \cite{Copeland:1997ug}.}. As the amplitude of
field perturbations at Hubble exit are of order $H$, we require the
super-Hubble perturbations to grow at the same rate to maintain a
scale-invariant spectrum. Ref.~{\cite{KW}} studied this instability
in a phase space analysis. The multi-field scaling solution was
shown to be a saddle point in field space and the late-time
attractor is the old ekpyrotic collapse dominated by a single-field.
A tachyonic instability drives the scaling solution towards the
late-time attractor (see also \cite{Lehners:2007ac,
Buchbinder:2007ad}).

In Ref.~{\cite{KW}}, we pointed out that the natural turning point
in the field space trajectory due to the instability of the scaling
solution might itself offer the possibility of converting the scale
invariant spectrum of isocurvature field perturbations into a scale
invariant spectrum of curvature perturbations. In this paper, we
confirm this expectation by explicitly solving the evolution
equations for perturbations in a two field model. We find that the
ratio between curvature perturbations and isocurvature perturbations
at the final old ekpyrotic phase is solely determined by the ratio
of exponents of the two exponential potentials,
% DW2
and the amplitude is set by the Hubble rate at the transition time.

\section{Homogeneous field dynamics}

We first review the background dynamics of the fields. During the
ekpyrotic collapse the contraction of the universe is assumed to be
described by a 4D Friedmann equation in the Einstein frame with
scalar fields with negative exponential potentials
\begin{equation}
 3 H^2 = V + \frac12 \dot\phi_i^2 \,,
\end{equation}
where
\be
 \label{Vi}
 V = - \sum_i V_i e^{-c_i\phi_i} \,,
\ee
and we take $V_i>0$
and set $8\pi G$ equal to unity.

The authors of \cite{Lehners:2007ac} found a scaling solution (previously
studied in \cite{Finelli:2002we,Guo:2003eu}) in which both fields roll
down their potential as the universe approaches a big crunch
singularity.
In this ekpyrotic scaling collapse we find
% DW3
a power-law solution for the scale factor
 \be
 \label{scalingp}
 a \propto (-t)^p \,, \quad {\rm where}\ p = \sum_i \frac{2}{c_i^2} < \frac13 \,,
\ee
where
\be
 \label{backgroundij}
 \frac{\dot\phi_i^2}{\dot\phi_j^2}
 = \frac{-V_i e^{-c_i\phi_i}}{-V_j e^{-c_j\phi_j}}
 = \frac{c_j^2}{c_i^2} \,.
\ee As we will see in the next section, it is possible to generate
scale-invariant isocurvature perturbations around this background.
However, the ekpyrotic scaling solution (\ref{backgroundij}) is
unstable.

In addition to the scaling solution we have fixed points
corresponding to any one of the original fields $\phi_i$ dominating
the energy density where the other fields have negligible energy
density. These correspond to the original ekpyrotic power-law
solutions where
\be
 a \propto (-t)^{p_i} \,, \quad {\rm where}\
 p_i = \frac{2}{c_i^2} \,,
\ee
for $c_i^2>6$. We find that any of these single field dominated
solutions is a stable local attractor at late times during
collapse.

In Ref.~\cite{KW}, the ekpyrotic scaling solution
(\ref{backgroundij}) was shown to be a saddle point in the phase
space. We briefly review the phase space analysis. Introducing phase
space variables \cite{Copeland:1997et,Heard:2002dr,Guo:2003eu}
\begin{eqnarray}
x_i &=& \frac{\dot{\phi}_i}{\sqrt{6} H}, \\
y_i &=& \frac{\sqrt{V_i e^{-c_i \phi_i}}}{\sqrt{3} H}.
\end{eqnarray}
the first order evolution equations for the phase space variables are given by
\begin{eqnarray}
\frac{d x_i}{d N} &=& -3 x_i (1- \sum_j x_j^2) - c_i \sqrt{\frac{3}{2}} y_i^2, \\
\frac{d y_i}{d N} &=& y_i \left(3 \sum x_j^2  - c_i \sqrt{\frac{3}{2}} x_i \right),
\end{eqnarray}
where $N = \log a$.
The Friedmann equation gives a constraint
\begin{equation}
\sum_j x_j^2 - \sum_j y_j^2 =1.
\end{equation}
There are $(n+2)$ fixed points of the system where $dx_i/dN = dy_i/dN
=0$.
\begin{eqnarray}
A : &&\;\;\; \sum_j x_j^2 =1, \quad y_j =0.\\
B_i : &&\;\;\; x_i =\frac{c_i}{\sqrt{6}}, \;\;\;
y_i=-\sqrt{\frac{c_i^2}{6}-1},
 \;\;\;
x_j=y_j=0, \;\; (\mbox{for} \;\;\; j \neq i),\\
B : &&\;\;\; x_j = \frac{\sqrt{6}}{3 p} \frac{1}{c_j}, \;\;\; y_j =
-\sqrt{\frac{2}{c_j^2 p} \left(\frac{1}{3p}-1 \right) }.
\end{eqnarray}
In this paper, we focus on the fixed points $B$ and $B_i$ assuming
$c_i^2 >6$
% DW
and $\sum c_i^{-2}<1/6$.
The linearized analysis shows that the multi-field scaling solution,
$B$, always has one unstable mode. On the other hand, the single
field dominated fixed points, $B_i$, are always stable.

{}From now on we concentrate our attention on two fields case. Then
we have three fixed points $B, B_1$ and $B_2$. It is interesting to
note that in the ($x_1$, $x_2$) plane, the fixed points $B$, $B_1$
and $B_2$ are connected by a straight line, which is given by
\begin{equation}
c_2 x_1 + c_1 x_2 = \frac{c_1 c_2}{\sqrt{6}}.
\label{instability}
\end{equation}
The eigenvector associated with the unstable mode around the scaling
solution $B$ lies in the same direction as the line
(\ref{instability}). Thus this is an attractor trajectory, which all
solutions near $B$ approach. Fig.~1 shows numerical solutions for
the evolution of $x_1$. Initial positions in the phase space are
perturbed away from $B$ along the line (\ref{instability}). The
solutions go to $B_1$ or $B_2$, depending on the initial position in
the phase space.

% DW FOLLOWING PARAGRAPH RE-WRITTEN
An important observation is that, as we follow phase space
trajectories during the transition from the scaling solution $B$ to
the attractor solutions $B_1$ or $B_2$, the solutions obey the
relation Eq.~(\ref{instability}), even far away from the saddle
point, $B$.
Using the Friedmann equation and the field equations, we can show
that
\begin{equation}
\left(H - \frac{\dot{\phi}_1}{c_1} - \frac{\dot{\phi}_2}{c_2}
\right)^{\cdot} + 3 H \left(H - \frac{\dot{\phi}_1}{c_1} -
\frac{\dot{\phi}_2}{c_2} \right)=0 \, ,
\end{equation}
% DW
and hence
\begin{equation}
H - \frac{\dot{\phi}_1}{c_1} - \frac{\dot{\phi}_2}{c_2} =
\frac{C}{a^3} \,,
\end{equation}
where $C$ is an integration constant.
In terms of $\phi_1$ and $\phi_2$, Eq.~(\ref{instability}) can be
rewritten as
\begin{equation}
\frac{\dot{\phi}_1}{c_1} + \frac{\dot{\phi}_2}{c_2} = H.
\label{instabilityf}
\end{equation}
and hence we see that for trajectories starting from point $B$ we
have $C=0$, which is the late time attractor.
Thus we see that Eq.~(\ref{instabilityf}) holds even during the
transition caused by the tachyonic instability from point $B$ to
$B_1$ or $B_2$ and in the final single-field dominated phase. This
fact will be important when we study perturbations.

\begin{figure}[http]
 \begin{center}
\includegraphics[width=17cm]{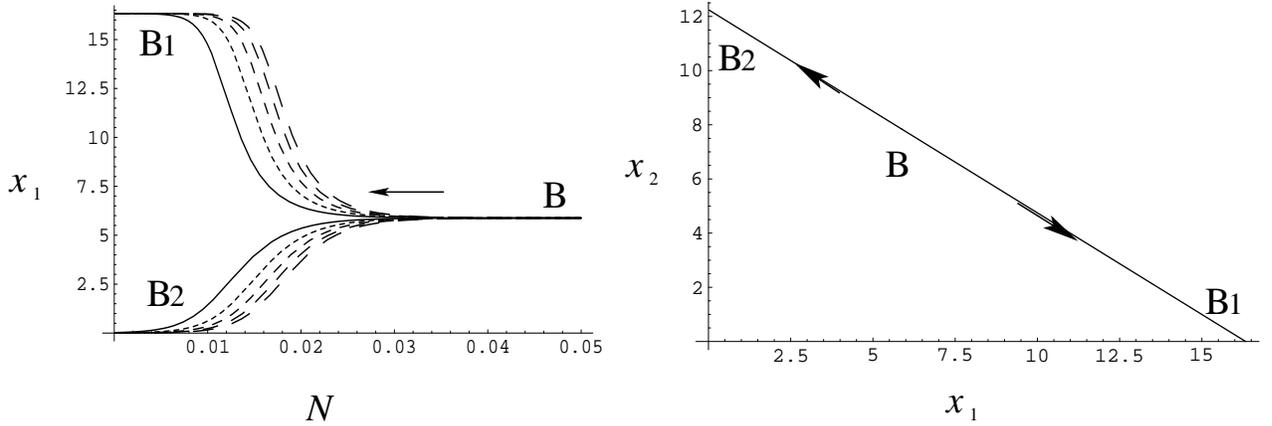}
\caption[]{Left: Numerical solutions for $x_1(N)$. The horizontal
axis is $N =\log a$ and we take $c_1 = 40$ and $c_2 = 30$. The
initial time is $N=0.05$. Note that $N$ decreases towards the future
in a collapsing universe. Right: The corresponding phase space
trajectories in $(x_1,x_2)$ plane.}
\end{center}
\end{figure}

We will study the behaviour of perturbations during the transition
in the next section.

\section{Generation of quantum fluctuations}

In this section, we consider inhomogeneous linear perturbations
around the background solution. We consider the scalar field
perturbations on spatially flat
% DW2
hypersurfaces.
Then the scalar field perturbations are given by
% DW3
\cite{Sasaki:1995aw,Taruya:1997iv,Gordon:2000hv,Hwang:2000jh}
\begin{equation}
\ddot{\delta \phi}_i + 3 H \dot{\delta \phi}_i + \frac{k^2}{a^2} \delta \phi_i
-c_i^2 V_i \exp(-c_i \phi_i)\delta \phi_i -\sum_j \frac{1}{a^3} \left(\frac{a^3}{H} \dot{\phi}_i
\dot{\phi}_j \right)^{\cdot} \delta \phi_j =0.
\end{equation}
We can decompose the perturbations into the instantaneous adiabatic
and entropy field perturbations as follows \cite{Gordon:2000hv}:
\begin{equation}
 \label{deltar}
\delta r = \frac{\dot{\phi}_1 \delta \phi_1 + \dot{\phi}_2 \delta \phi_2}{
\sqrt{\dot{\phi}_1^2 + \dot{\phi}_2^2}}, \quad
 \label{deltas}
\delta s = \frac{\dot{\phi}_2\delta \phi_1 - \dot{\phi}_1 \delta
\phi_2}{ \sqrt{\dot{\phi}_1^2 + \dot{\phi}_2^2}}.
\end{equation}
The adiabatic field perturbation $\delta r$ is the component of the
two-field perturbation along the direction of the background fields'
evolution while the entropy perturbation $\delta s$ represents
fluctuations orthogonal to the background classical trajectory.
% DW
The adiabatic field perturbation leads to a perturbation in the
comoving curavture perturbation:
\begin{equation}
 \label{Rc}
 {\cal R}_c = \frac{H\delta r}{\dot{r}} \,,
 \end{equation}
whereas the entropy field perturbations correspond to isocurvature
perturbations.

Their evolution equations are given by \cite{Gordon:2000hv}
\begin{equation}
\label{evolr}
\ddot{\delta r} + 3 H \dot{\delta  r} +
\frac{k^2}{a^2} \delta r +
\left[ V_{,rr}- \dot{\theta}^2 -
\frac{1}{a^3} \left(\frac{a^3 \dot{r}^2}{H} \right)^{\cdot} \right] \delta r
= 2 \dot{\theta} \dot{\delta s} + 2 \left[ \ddot{\theta} -
\left(\frac{V_{,r}}{\dot{r}} + \frac{\dot{H}}{H} \right) \dot{\theta} \right] \delta s,
\end{equation}
\begin{equation}
 \label{evols}
\ddot{\delta s} + 3 H \dot{\delta s} +
\frac{k^2}{a^2} \delta s + (V_{,ss} - \dot{\theta}^2) \delta s
= -2 \frac{\dot{\theta}}{\dot{r}}
 \left[
 \dot{r} \dot{\delta r} - \left(\frac{\dot{r}^3}{2H} + \ddot{r}\right) \delta r
 \right],
\end{equation}
where the angle $\theta$ is defined as
\begin{equation}
\cos \theta = \frac{\dot{\phi_2}}{\sqrt{\dot{\phi}_1^2 + \dot{\phi}_2^2}},
\quad
\sin \theta = \frac{\dot{\phi_1}}{\sqrt{\dot{\phi}_1^2 + \dot{\phi}_2^2}},
\end{equation}
such that
\begin{eqnarray}
\dot{r} &=& (\cos \theta) \dot{\phi}_2 + (\sin \theta) \dot{\phi}_1, \\
\dot{\theta} &=&-\frac{V_{,s}}{\dot{r}}
 \,,
\end{eqnarray}
and
\begin{eqnarray}
V_{,r} &=& (\sin \theta) c_1 V_1 \exp(-c_1 \phi_1)+(\cos \theta) c_2 V_2\exp(-c_2 \phi_2), \\
V_{,s} &=& (\cos \theta) c_1 V_1 \exp(-c_1 \phi_1) - (\sin \theta) c_2 V_2 \exp(-c_2 \phi_2),\\
V_{,rr}&=&- (\sin \theta)^2 c_1^2 V_1 \exp(-c_1 \phi_1)
-(\cos \theta)^2 c_2^2 V_2 \exp(-c_2 \phi_2), \\
V_{,ss}&=&- (\sin \theta)^2 c_2^2 V_2 \exp(-c_2 \phi_2)
-(\cos \theta)^2 c_1^2 V_1 \exp(-c_1 \phi_1).
\end{eqnarray}

For the multi-field scaling solution, $B$, we have
\begin{equation}
\theta = \arctan\frac{c_2}{c_1} \,,
\end{equation}
and for the single field scaling solutions we have
\begin{equation}
\theta =\frac{\pi}{2}  \;\;\; (B_1) \,, \quad \theta = 0 \;\;\;
(B_2) \,.
\end{equation}
Thus we have $\theta=$constant for the fixed points and the
adiabatic and the entropy fields are decoupled. This allows us to
quantise the independent fluctuations in the two fields.

For the multi-field scaling solution $B$, the spectrum of quantum
fluctuations of the entropy field is given on large scales ($k\ll
aH$) by
 \be
 \label{Pdeltachi}
 {\cal P}_{\delta s} \equiv \frac{k^3}{2\pi^2} |\delta s^2|
 = C_\nu^2 \frac{k^2}{a^2} (-k\tau)^{1-2\nu} \,,
\ee
% DW
where $\tau<0$ is conformal time,
and
\begin{equation}
\nu^2 = \frac94 - \frac{3 \epsilon}{(\epsilon-1)^2},
\quad \epsilon \equiv -\dot{H}/H^2 =1/p,
\end{equation}
and $C_\nu=2^{\nu-3/2}\Gamma(\nu)/\pi^{3/2}$ \cite{Finelli:2002we}.
The spectral tilt is given by
\be
 \Delta n_{\delta\chi} \simeq \frac{2}{\epsilon} \,,
\ee
to leading order in a fast-roll expansion
($\epsilon\gg1$) \cite{Lehners:2007ac, Buchbinder:2007ad,
Creminelli:2007aq}.
In this limit, the spectrum (\ref{Pdeltachi}) can be written as
\begin{equation}
 \label{Ps}
{\cal P}_{\delta s}^{1/2}= \epsilon \left|\frac{H}{2 \pi} \right| \,.
\end{equation}
Note that $|H|$ is rapidly increasing and thus $\delta s$ is also
growing on super-Hubble scales due to the tachyonic instability.
This instability is essential in order to realize the scale invariance of the
spectrum (\ref{Ps}). The amplitude of field perturbations at Hubble
exit is of order $H$ and thus we require the super-Hubble
perturbations to grow at the same rate in order to maintain a
scale-invariant spectrum.

% DW moved here from the conclusions as this is a minor comment
Note that in the simplest model, the spectrum is slightly blue
\cite{Lehners:2007ac, Buchbinder:2007ad, Creminelli:2007aq}.  However,
any deviations from an exponential potential for adiabatic field can
introduce the corrections to the spectral tilt and thus it becomes
model dependent \cite{Lehners:2007ac, Buchbinder:2007ad}.

% DW MORE DETAIL GIVEN ON ADIABATIC PERTURBATIONS
The spectrum of quantum fluctuations in the adiabatic field about
the scaling solution has the same power-law form on large scales
 \begin{equation}
 {\cal P}_{\delta r} = C_\mu^2 \frac{k^2}{a^2} (-k\tau)^{1-2\mu} \,,
 \end{equation}
where to $\mu\simeq 1/2$ to leading order in $1/\epsilon$. Thus the
adiabatic field perturbations become constant in the large scale
limit and the spectral tilt is given by
\begin{equation}
\Delta n_{\delta r} \simeq 2 \,.
\end{equation}
Thus we have
\begin{equation}
 \frac{{\cal P}_{\delta r}}{{\cal P}_{\delta s} } \propto
 (-k\tau)^2 \,,
\end{equation}
and hence in what follows we can neglect the adiabatic field
fluctuations in the large-scale limit.

By contrast, for the single field dominated scaling solutions, the
adiabatic and entropy field perturbations are both frozen on
super-Hubble scales:
\begin{equation}
\delta s \,, \, \delta r = \mbox{const.}
\end{equation}
These perturbations both have a steep blue spectrum if they cross the
horizon when the background solutions are described by the single
field dominated solution \cite{KW}.

\section{Generation of curvature perturbations}

Now let us consider the evolution of perturbations in a situation
where the classical solution starts from near the saddle point, $B$.
As emphasized in Ref.~\cite{KW} this requires an additional
preceding mechanism that drives the classical background solution to
the unstable saddle point throughout our observable part of the
universe. In this paper, we will not discuss the mechanism required
to bring the classical solution to the saddle point and we just
assume that the classical solution stays near the saddle point for
long enough to ensure that a scale-invariant spectrum of
isocurvature perturbations is generated over the relevant scales for
the observed large scale structure of our Universe.

Then the initial conditions for the adiabatic and the entropy field
perturbations can be set from the amplitude of quantum fluctuation as
described in the previous section:
%
% DW modulus sign replacing bracket
\begin{equation}
\delta r =0, \quad  \delta s = \epsilon \left|\frac{H}{2 \pi}
\right| \,,
 \label{initial}
\end{equation}
on sufficiently large scales and for $1/\epsilon \ll 1$.

Unless the spatially homogeneous background solution is located
exactly at the fixed point, the tachyonic instability drives the
background solution away from the multi-field scaling solution, $B$,
to one of the single field dominated solutions, $B_1$ or $B_2$,
depending on the initial conditions. During the transition $\theta$
is not constant and the adiabatic and entropy field perturbations
mix, so it is possible to generate perturbations in the adiabatic
field, and hence comoving curvature perturbations (\ref{Rc}), from
initial fluctuations in the entropy field.

We can solve the evolution equations (\ref{evolr}) and
(\ref{evols}) numerically for any given classical background.
Figure~2 shows the behaviour of $\delta r$ and $\delta s$. Due to the
coupling between $\delta r$ and $\delta s$ during the transition,
curvature perturbations are generated during the transition. On the
other hand, $\delta s$ shows a tachyonic instability according to
Eq.~(\ref{initial}) close to the scaling solution, but when the
background solution goes to $B_i$, the entropy field perturbation
becomes constant. The final amplitude of $\delta r$ depends on when
the transition from $B$ to $B_i$ occurs, but, interestingly, the final
ratio between $\delta r$ and $\delta s$ does not depend on the details of
the transition. We find that the ratio is determined solely by the
parameters $c_1$ and $c_2$ as
\begin{eqnarray}
\frac{\delta r}{\delta s } &=& \frac{c_1}{c_2}, \quad {\rm at}\ B_1,
\label{ratio1}
\\
\frac{\delta r}{\delta s} &=& -\frac{c_2}{c_1}, \quad {\rm at}\ B_2,
\label{ratio2}
\end{eqnarray}
as is shown in Fig.~2.
% DW comment added
We will explain later why such a simple result is found.

\begin{figure}[http]
 \begin{center}
\includegraphics[width=17cm]{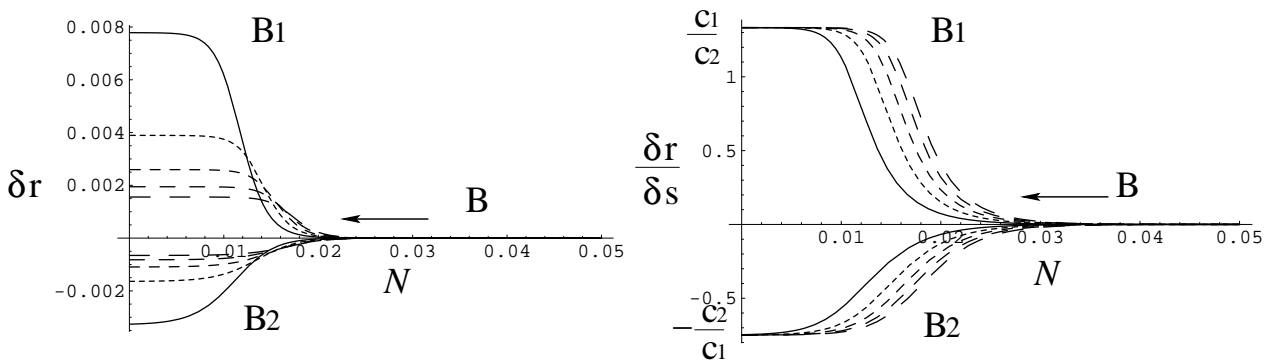}
\caption[]{Left: Solutions for $\delta r(N)$, using the same
parameters as in Figure~1. The corresponding background solutions
are shown in Figure~1. Right: The ratio between $\delta r$ and
$\delta s$. The ratio approaches a constant given by
Eqs.~(\ref{ratio1}) and (\ref{ratio2}). In this case $1.3333$ for
$B_1$ and $-0.75$ for $B_2$.}
\end{center}
\end{figure}

The resulting curvature perturbation on a comoving hypersurface in
the final single-field dominated phase is thus given by
\begin{eqnarray}
{\cal R}_c &=& \frac{H}{\dot{\phi}_1} \delta r = \frac{1}{c_1} \delta
r, \quad {\rm at}\ B_1, \\
{\cal R}_c &=& \frac{H}{\dot{\phi}_2} \delta r = \frac{1}{c_2} \delta
r, \quad {\rm at}\ B_2.
\end{eqnarray}

The equation for the entropy field
perturbation (\ref{evols}) can be rewritten as \cite{Gordon:2000hv}
\begin{equation}
\ddot{\delta s} + 3 H \dot{\delta s} + \left( \frac{k^2}{a^2} +
V_{ss} + 3 \dot{\theta}^2 \right) \delta s = 4
\frac{\dot{\theta}}{\dot{r}} \frac{k^2}{a^2} \Psi \,,
\label{entropyPsi}
\end{equation}
where $\Psi$ is the curvature perturbations in Newtonian gauge.
The change of the curvature perturbations is determined by \cite{Gordon:2000hv}
\begin{equation}
\dot{{\cal R}}_c = \frac{H}{\dot{H}} \frac{k^2}{a^2} \Psi
+ \frac{2 H}{\dot{r}} \dot{\theta} \delta s.
\label{curvaturePsi}
\end{equation}
On large scales, we can neglect $(k^2/a^2) \Psi$ and it is possible 
to reproduce the previous results from Eq.~(\ref{curvaturePsi}).

%In Refs.~\cite{Lehners:2007ac, Buchbinder:2007ad}, the curvature
%perturbation generated from the initial entropy perturbations are
%estimated by neglecting $(k^2/a^2) \Psi$ on large scales, as can be
%done during slow-roll inflation \cite{Gordon:2000hv}. However, this
%assumption is not necessarily justified in a collapsing universe
%where $\Psi$ may grow rapidly on large scales. Even in a single
%field inflation where $\delta s=0$, there are cases where ${\cal R}_c$
%can change its amplitude on super-horizon scales
%\cite{Leach:2001zf}. In fact, in our case, we checked
%that the contribution from the $\Psi$ terms in
%Eqs.~(\ref{entropyPsi}) and (\ref{curvaturePsi}) cannot be neglected
%during the transition from scaling solution to the single-field
%dominated solution.

Although the physical meaning of the instantaneous adiabatic and
entropy field perturbations is clear, the dynamics of the
perturbations during the transition are rather complicated in this
basis. We find it is much easier to work in terms of new variables
\cite{KW}
\be
 \varphi = \frac{c_2 \phi_1 + c_1 \phi_2}{\sqrt{c_1^2+c_2^2}} \,,
\quad \chi = \frac{c_1 \phi_1 - c_2 \phi_2}{\sqrt{c_1^2+c_2^2}} \,,
\label{sigmachi}
 \ee
corresponding to a fixed rotation in field space.
The potential Eq.~(\ref{Vi}) can then be simply re-written as
\cite{Malik:1998gy,Finelli:2002we,KW}
 \be
 \label{Vsigmachi}
 V = - U(\chi) \, e^{-c\varphi} \,,
\ee
where
\be
\frac{1}{c^2} = \sum_i \frac{1}{c_i^2} \,,
\label{c}
\ee
and the potential for the orthogonal field is given by
\be
- U(\chi) = - V_1\, e^{-(c_1/c_2)c\chi} - V_2\, e^{(c_2/c_1)c\chi} \,,
\ee
which has a maximum at
\be
\chi=\chi_0=\frac{1}{\sqrt{c_1^2+c_2^2}}
\ln \left(\frac{c_1^2 V_1}{c_2^2 V_2}\right).
\ee

The multi-field scaling solution corresponds to $\chi =\chi_0$,
while $\varphi$ is rolling down the exponential potential. The
potential for $\chi$ has a negative mass-squared around $\chi
=\chi_0$, and thus $\chi$ represents the instability direction. If the
initial condition for $\chi$ is slightly different from $\chi_0$ or
$\dot{\chi}$ is not zero, then $\chi$ starts rolling down the potential
and the solution approaches a single-field dominated solution.

Note that perturbations $\delta\varphi$ and $\delta\chi$ coincide with
the instantaeous adiabatic and entropy field perturbations
respectively, defined in Eq.~(\ref{deltar}), at
the scaling solution, $B$. Thus we can use the initial perturbations
(\ref{initial}) due to vacuum fluctuations about the scaling solution
previously calculated. However as we follow the evolution away from
this saddle point we can no longer identify $\varphi$ and $\chi$ with
the adiabatic and entropy perturbations. Nevertheless, the dynamics of
perturbations turns out to be much simpler using these fields.

In terms of $\varphi$ and $\chi$
the equations for perturbations are given by
\begin{eqnarray}
\ddot{\delta \varphi} + 3 H \dot{\delta \varphi} +\frac{k^2}{a^2}
\delta \varphi +M_{\varphi \varphi} \delta \varphi + M_{\varphi \chi} \delta \chi=0, \\
\ddot{\delta \chi} + 3 H \dot{\delta \chi} +\frac{k^2}{a^2}
\delta \chi +M_{\chi \chi} \delta \chi + M_{\varphi \chi} \delta \varphi=0,
\end{eqnarray}
where
\begin{eqnarray}
M_{\varphi \varphi} &=& V_{,\varphi \varphi} - \frac{1}{a^3}
\left(\frac{a^3}{H} \dot{\varphi}^2 \right)^{\cdot}, \\
M_{\varphi \chi} &=&  V_{,\varphi \chi} - \frac{1}{a^3}
\left(\frac{a^3}{H} \dot{\varphi} \dot{\chi} \right)^{\cdot},\\
M_{\chi \chi} &=&  V_{,\chi \chi} - \frac{1}{a^3}
\left(\frac{a^3}{H} \dot{\chi}^2 \right)^{\cdot}.
\end{eqnarray}

% DW next equation moved here
A key observation is that the phase space trajectory of the background
fields during the transition, Eq.~(\ref{instabilityf}), can be re-written as
\begin{equation}
\frac{\dot{\varphi}}{H} = c.
\label{sigmacon}
\end{equation}
Thus even away from the multi-field scaling solution, $\varphi$
obeys a scaling relation.
We can then show that two of the effective mass terms become
\begin{eqnarray}
M_{\varphi \varphi} &=& V_{,\varphi \varphi} + c V_{,\varphi} = 0 \,, \\
M_{\varphi \chi} &=& V_{,\varphi \chi} + c V_{,\chi} = 0 \,,
\end{eqnarray}
independently of the form of $U(\chi)$ since $V \propto \exp(-c \varphi)$.

Thus on large scales $\delta \varphi$ is constant. As we take $\delta
\varphi =0$ as our initial condition, this remains so even during the
transition and in the final single field dominated phase. Thus from
Eq.~(\ref{sigmachi}) we have a relation between $\delta \phi_1$ and
$\delta \phi_2$,
\begin{equation}
c_2 \delta \phi_1 + c_1 \delta \phi_2 = 0,
\end{equation}
and $\delta \phi_1$ and $\delta \phi_2$ are given by
\begin{eqnarray}
\delta \phi_1 &=& \frac{c_1}{\sqrt{c_1^2+c_2^2}} \delta \chi, \\
\delta \phi_2 &=& -\frac{c_2}{\sqrt{c_1^2+c_2^2}} \delta \chi.
\end{eqnarray}
In the single field dominated solutions, $\varphi$ is
no longer the adiabatic field. The adiabatic field
is simply $\phi_1$ (or $\phi_2$) in $B_1$ (or $B_2$) and the entropy
field is $-\phi_2$ (or $\phi_1$) in $B_1$ (or $B_2$). Thus at the
late-time attractor we have
\begin{eqnarray}
\delta r &=&  \delta \phi_1, \quad
\delta s = -\delta \phi_2, \quad {\rm at}\ B_1,\\
\delta r &=& \delta \phi_2, \quad
\delta s =  \delta \phi_1, \quad {\rm at}\ B_2.
\end{eqnarray}
Thus the ratio between $\delta r$ and $\delta s$ is
determined by the ratio between $\delta \phi_1$ and
$\delta \phi_2$ and we can easily find the ratio $\delta r/\delta s$
as given in Eqs.(\ref{ratio1}) and (\ref{ratio2}).

The amplitude of the curvature perturbations can be estimated
from $\delta \chi$. In the initial stage, $\delta \chi$
coincides with the entropy field perturbations $\delta s$, and thus
its initial amplitude on super-Hubble scales is given by
\begin{equation}
\delta \chi = \frac{c^2}{2} \left|\frac{H}{2 \pi} \right|.
\end{equation}
where we used $\epsilon = c^2/2$ in Eq.~(\ref{initial}) and
$c^2$ is given by Eq.~(\ref{c}).
After the transition, $\delta \chi$ becomes massless and
the amplitude becomes frozen on large scales. In the single field dominated
solutions, the comoving curvature perturbation ${\cal R}_c$ is given
by
\begin{eqnarray}
\vert {\cal R}_c \vert &=& \frac{1}{\sqrt{c_1^2+c_2^2}} \delta \chi.
\end{eqnarray}
Assuming the transition occurs suddenly, the final amplitude of the
comoving curvature perturbation is given by
\begin{equation}
\vert {\cal R}_c \vert = \frac{c^2}{2 \sqrt{c_1^2+c_2^2}}
\left|\frac{H}{2 \pi} \right|_{T},
\label{rc}
\end{equation}
where the subscript T means that the quantity is evaluated
at the transition time.
On the other hand the amplitude
of the entropy field perturbation is given by
\begin{eqnarray}
\delta s &=& \frac{c_2 c^2}{2\sqrt{c_1^2 +c_2^2}}
\left|\frac{H}{2 \pi} \right|_{T}, \quad {\rm at}\ B_1,
\label{deltas1}
\\
\delta s &=& \frac{c_1 c^2}{2\sqrt{c_1^2 +c_2^2}}
\left|\frac{H}{2 \pi} \right|_{T}, \quad {\rm at}\ B_2.
\label{deltas2}
\end{eqnarray}
{}From the numerical solutions for $\delta r$ and $\delta s$,
we can reconstruct $H_T$. We confirm that $H_T$ constructed
in this way agrees with the Hubble scale at the transition
as is shown in Figure~3.

\begin{figure}[http]
 \begin{center}
\includegraphics[width=17cm]{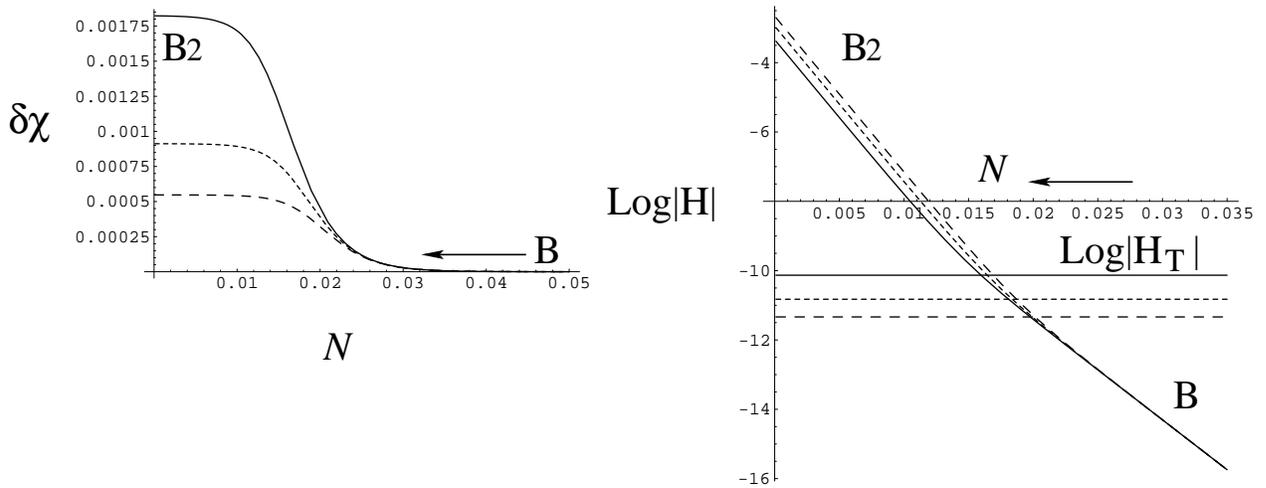}
\caption[]{Left: Solutions for $\delta \chi(N)$. We used the
same parameters as Figures~1 and~2. Right: Solutions for
$\log \vert H \vert $ with three different background solutions.
We also show $\log \vert H_T \vert $ that is determined from the
numerical solutions for $\delta \chi$.}
\end{center}
\end{figure}

\section{Conclusion}

In this paper we have studied the generation of curvature
perturbations during an ekpyrotic collapse with multiple fields.  We
must assume that the classical background solution starts from a state
very close to a saddle point in the phase space that corresponds to
the multi-field ekpyrotic scaling solution. If the solution stays at
this fixed point for long enough, a scale-invariant spectrum of
isocurvature perturbations is generated over the range of scales that
is relevant for large scale structure in the Universe.  Even if the
background solution deviates only slightly from the multi-field
scaling solution initially, the tachyonic instability eventually
drives the solution to the old ekpyrotic collapse dominated by a
single field. During this transition, the initial isocurvature field
perturbations generate a scale-invariant spectrum of comoving
curvature perturbations.

First we studied the perturbations by decomposing them into the
instantaneous adiabatic and the entropy field perturbations. These
fields are decoupled at the fixed points. The adiabatic field
perturbations are effectively massless around both the multi-field
scaling solutions and the single field dominated solution, so become
constant on large scales. On the other hand, the entropy field
perturbations have a tachyonic mass around the multi-field scaling
solution, where they grow like $\delta s \propto H$ on large scales,
but they are effectively massless around the single field dominated
solution. We set initial conditions for these perturbations from
quantum fluctuations about the multi-field scaling solution.  As the
adiabatic field perturbations have a blue spectrum they can be
neglected compared with the entropy field perturbations on large
scales. However, during the transition to the single-field attractor,
adiabatic field perturbations are generated from the entropy field
perturbations.
% DW2 some repetitive sentences omitted in the conclusions
%The final amplitude of the perturbations depends on when the
%transition occurs from the multi-field scaling solution to the
%single field dominated solution. However, we found that the ratio
%between the final adiabatic and entropy field perturbations is
%solely determined by the ratio of exponents of the potentials as in
%Eqs.~(\ref{ratio1}) and (\ref{ratio2}).

It turns out to be more convenient to use new fields $\varphi$ and
$\chi$ defined in Eqs.~(\ref{sigmachi}) to follow the perturbations
through the transition.  These fields coincide with the adiabatic
and entropy fields around the multi-field scaling solution but not
during the transition or at the final single field dominated phase.
In terms of $\varphi$ and $\chi$, the potential is given by a
product of the potential for $\chi$, $U(\chi)$, and an exponential
potential for $\varphi$.  $U(\chi)$ has an extremum which
corresponds to the multi-field scaling solution. A crucial
observation is that even during the transition and in the final
phase, $\varphi$ satisfies the scaling relation
Eq.~(\ref{sigmacon}).  We can then show that the field perturbations
for $\varphi$ and $\chi$ are always decoupled. Since $\delta
\varphi=0$ on large scales during the ekpyrotic scaling solution,
this remains so. This determines the ratio between the adiabatic and
entropy field perturbations at the final phase. On the other hand,
$\delta \chi$ grows during the scaling solution and then becomes
constant during the single field dominated solution.
% DW2
%The final amplitude of the comoving curvature perturbation is then
%determined by the value of the $\delta\chi$ at the transition.

Our final results are Eqs.~(\ref{rc}), (\ref{deltas1}) and
(\ref{deltas2}) for the amplitude of comoving curvature and
isocurvature field perturbations during the single field ekpyrotic
collapse phase. The amplitude of the comoving curvature perturbation
is determined by the Hubble scale at the transition.

We still need to convert the ekpyrotic collapse to expansion (see,
for instance, \cite{Buchbinder:2007ad, Creminelli:2007aq}) and see
how this curvature perturbation is matched to that in an expanding
universe. For a regular bounce the comoving curvature perturbation
is conserved for adiabatic perturbations and thus Eq.~(\ref{rc}) is
directly related to the amplitude of the observed primordial density
perturbation.  If the radiation and matter content in today's
universe
% DW3
comes solely from the single field that dominates the final
ekpyrotic phase, we will have no isocurvaure perturbations in an
expanding universe.

% DW new paragraph incorporating comments on tensor spectrum
It is interesting to compare this multi-field model with a single
field model that gives a scale-invariant spectrum of comoving
curvature perturbations during collapse
\cite{Wands:1998yp,Finelli:2001sr,Allen:2004vz}. The single-field
model requires the correct exponent for a relatively flat and
positive exponential potential in order to obtain $a\propto
|t|^{2/3}$, whereas the ekyprotic model only requires sufficiently
steep, negative exponential potentials to obtain $a\propto
|t|^{1/\epsilon}$ with $\epsilon\gg1$. On the other hand both models
require fine-tuned initial conditions as it is the existence of an
instability that gives rise to the scale-invariant perturbation
spectrum during collapse.  In both models the amplitude of tensor
perturbations is determined by the Hubble scale when the
perturbations leave the horizon as tensor perturbations are then
frozen on super-horizon scales.  In the single field model the
tensor metric perturbations thus acquire the same almost
scale-invariant spectrum \cite{Starobinsky:1979ty} as the
% DW2
comoving
curvature perturbation, with a similar amplitude, giving rise to a
dangerously large tensor-scalar ratio which severely constrains the
model \cite{Allen:2004vz}. But in the ekpyrotic scaling solution the
tensor perturbations have a steep blue spectrum and are completely
negligible at scales relevant for the cosmic microwave anisotropies.

In summary, a simple ekpyrotic model with two steep, negative
exponential potentials is capable of generating a scale-invariant
spectrum of comoving curvature perturbations. A key ingredient is the
instability of the multi-field scaling solution
\cite{Finelli:2002we,Lehners:2007ac, Buchbinder:2007ad,KW}. This
instability generates a scale-invariant spectrum of isocurvature field
perturbations from vacuum fluctuations about the scaling solution, and
converts them to a scale-invariant spectrum of comoving curvature
perturbations. We should emphasize that this conversion occurs
automatically due to the dynamics of the fields in our simple model
and does not require any change in the shape of the potential or any
additional dynamics \cite{Lehners:2007ac,Buchbinder:2007ad,
Creminelli:2007aq}.

Thus ekpyrotic collapse with multiple fields can generate a
scale-invariant spectrum for curvature perturbations in several
different ways (see also \cite{Tolley:2007nq} for a different idea).
In all these models, we need some preceding phase that initially
drives the classical background solution to the unstable multi-field
scaling solution throughout our observable region of space. This
problem cannot be solved within the simplest model with multiple
exponential potentials
% DW2
that we considered in this paper and we would need to appeal to a
more ambitious framework for the model such as the cyclic scenario
\cite{Steinhardt:2001st,Steinhardt:2002ih} to address this problem.

\section*{Acknowledgments}
We would like to thank J-L.~Lehners and P.J.~Steinhardt for 
discussions. KK and DW are supported by STFC. SM is grateful to the ICG,
Portsmouth, for their hospitality when this work was initiated. SM
is supported in part by the Japan Society for Promotion of Science
(JSPS) Research Fellowship.

\end{document}